# Family Matters

Ferruccio Feruglio[1] *

[1]INFN, Sezione di Padova, Via Marzolo 8, I-35131 Padua, Italy

**Abstract**

Quarks and leptons, the fundamental building blocks of the subatomic world, manifest in three families - replicas with identical quantum numbers that differ only in their masses. After revisiting the key milestones that led to the discovery of this peculiar structure, a non-technical overview is presented of the main attempts to explain its origin and trace it back to an as-yet-unknown fundamental principle.

*E-mail: `feruglio@pd.infn.it`

# 1 A misplaced piece

Like in a crossword puzzle where the final word doesn't quite fit, forcing a revision of many definitions, the particle discovered in 1937 by Neddermeyer and Anderson was destined to profoundly revolutionize physics. Yet, nothing seemed capable of challenging a well-established framework: ordinary matter was composed of protons (Rutherford, 1919), neutrons (Chadwick, 1932), and electrons (Thomson, 1897). To these foundations was added Pauli's hypothesis of the neutrino, while in 1934 Fermi developed a theory of beta decay, a cornerstone of the current electroweak theory. With Dirac's 1928 equation, antimatter acquired a theoretical form, soon confirmed by its actual existence.

The particle discovered by Neddermeyer and Anderson in cosmic radiation stood out for its high penetrating power [1]. It was absorbed more readily by the atmosphere than by an equivalent layer of solid material, due to its in-flight decay with a mean lifetime of a few microseconds. It could carry either a positive or negative charge and had a mass intermediate between that of the electron and the proton. In 1935, Yukawa had theorized the existence of "mesotrons", mediators of nuclear forces, approximately 200 times more massive than the electron, with positive, negative, or neutral charge. The identification of Neddermeyer and Anderson's particle with Yukawa's mesotron thus seemed natural and promised to unravel the mystery of nuclear forces. Doubts regarding this matter were temporarily set aside due to the looming global conflict.

It was during the bombings of World War II that an experiment was conceived and conducted in Rome, one that would unveil the true nature of the newly discovered particle. In February 1947, following a series of experiments, Pancini, Piccioni, and Conversi published a paper on the decay of the presumed mesotrons at rest, achieved using a carbon absorber. According to theoretical predictions, only the decays of positively charged particles should have been observed, while negatively charged particles would be captured by the atoms of the material and absorbed by the nuclei, without sufficient time to decay. However, the experiment revealed an unexpected result: in carbon, negatively charged particles decayed just like their positively charged counterparts, in clear contradiction to Yukawa's theory.

I have a vivid memory of Milla Baldo-Ceolin's Advanced Physics lectures in Padova, where she often emphasized the significance of this experiment. Luis Alvarez, in his 1968 Nobel Lecture, remarked: "In my personal opinion, modern particle physics began in the closing days of World War II, when a group of young Italian physicists - Conversi, Pancini, Piccioni - hiding in Rome from German occupying forces, undertook an experiment of extraordinary importance."

The situation was finally explained in October 1947, thanks to Lattes, Powell, and Occhialini. By studying cosmic rays with nuclear emulsions exposed at high altitudes, the three researchers identified Yukawa's mesotron, now known as the pion, and its decay into the particle discovered by Neddermeyer and Anderson: the muon. While the pion emerged as the key to describing nuclear interactions, the role of the muon remained entirely obscure.

# 2 Bruno Pontecorvo makes the difference

In June of the same year, Bruno Pontecorvo made a crucial observation: the absorption rate of muons by nuclei was analogous to the nuclear capture rate of electrons, once the different phase spaces and Bohr radii were taken into account. This insight, remarkable given that the two rates differ by about eleven orders of magnitude, carried profound implications [2]. Pontecorvo suggested that the muon was a sort of "older brother" to the electron, differing only in its larger mass. Regarding interaction properties, the electron and muon appeared indistinguishable. But why did the electron have a massive counterpart? What exactly were the properties of this counterpart? What role did it play in fundamental interactions? Did it exist on its own, or were there other partners involved?

It was soon discovered that the muon decayed into an electron and two neutrinos, possessing spin 1/2. Decays into final states without neutrinos, such as $\mu \to e\gamma$, $\mu \to 3e$, or $\mu + (A, Z) \to e + (A, Z)$, however, were not observed, despite the lack of known selection rules at the time prohibiting them. The suppression of these decays required an explanation.

The hypothesis that weak interactions were mediated by a heavy boson, coupled exclusively to charged currents, provided a partial answer. It explained the absence of direct interactions between four charged fermions or four neutral fermions but did not entirely rule out unobserved processes, which could have occurred through the emission and absorption of the intermediate boson. For instance, the decay $\mu \to e\gamma$ was predicted to occur via a second-order weak transition involving the exchange of an intermediate boson and a neutrino (Fig. 1). Estimates of the probability of this transition were inaccurate due to the presence of divergent quantities. However, by the late 1950s, experimental limits exceeded the most conservative theoretical predictions by more than an order of magnitude [3]. In

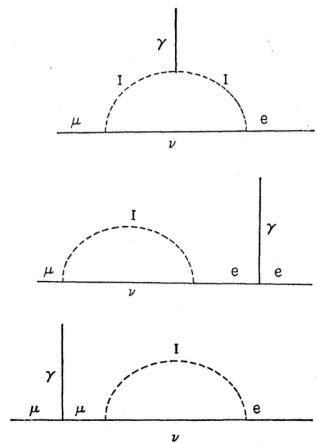

**Figure 1:** Transition $\mu \to e\gamma$ from the exchange of a neutrino and an intermediate boson I.

June 1959, Bruno Pontecorvo observed that, in the presence of two distinct neutrinos - one associated with the muon and one with the electron - the transition $\mu \to e\gamma$ would not be possible. The diagram involving the exchange of an intermediate boson and a single neutrino could not be realized, as the boson would interact exclusively with the electron and the electron neutrino, or with the muon and the muonic neutrino. At the time, neutrinos were thought to be massless, making the idea of mixing between two distinct species of neutrinos inconceivable. In the same work, Pontecorvo proposed a series of experimental tests to verify his hypothesis [4].

Once again, Pontecorvo's intuition proved extraordinarily innovative. In fact, he introduced a conserved quantum number, now known as the family lepton number. After many years, we now know that the family lepton number is violated in neutrino oscillations, but it continues to serve as an effective selection rule in transitions involving charged leptons.

The exact cancellation proposed by Pontecorvo in the transition amplitudes for processes such as $\mu \to e\gamma$ and $\mu \to 3e$ anticipated, in some aspects, the partial cancellation observed in flavor-changing neutral current transitions, later explained by the GIM mechanism. Moreover, the absence of decays into final states without neutrinos implies the necessity of at least two families of neutrinos. This led to an interpretation of the muon not as a "desert flower" but as part of a larger system, accompanied by at least one other particle, which was finally observed in 1962 by Leon Max Lederman, Melvin Schwartz, and Jack Steinberger.

## 3 The hadrons proliferate

If the electron had found a "big brother" and the leptonic family had expanded to include two neutrinos, the extension to a second generation in the hadronic sector took much longer, following a far more complex path. The 1950s were marked by a flurry of new discoveries, so much so that Willis Lamb, Nobel laureate in Physics in 1955, ironically remarked: "The discovery of a new elementary particle once deserved the Nobel Prize; now it should be punished with a $10,000 fine."

The new hadronic particles were initially identified through the study of cosmic rays. Later, the advent of particle accelerators allowed for their systematic analysis. The new hadrons were defined as "strange" due to a peculiar feature: their production timescale was much shorter than their decay timescale. To explain this property, in 1953, Gell-Mann and Nishijima introduced a new quantum number, "strangeness," which was conserved in strong interactions but violated in weak ones. Strange particles could not decay into "ordinary" hadrons (such as protons, neutrons, and pions) via strong interactions; their relative stability was due to the weak interaction, which governed their decay. Moreover, the production of strange particles from ordinary hadrons via strong interactions required that they be generated in pairs [5].

A first classification of hadrons had to wait until 1961, with the introduction of Murray Gell-Mann's Eightfold Way, followed in 1964 by the interpretation of all hadrons as composite states made up of elementary constituents, the quarks (Gell-Mann and Zweig). Three types of quarks, or "flavors", were hypothesized: *up*, *down*, and *strange*, with electric charges of +2/3, -1/3, and -1/3, respectively. Each baryon was interpreted as a system composed of three quarks, each antibaryon of three antiquarks, and each meson of a quark and an antiquark. Since, neglecting quark masses, strong interactions do not distinguish among the three flavors, the system exhibits symmetry under the SU(3) group. This implies that all observed hadrons organize themselves into SU(3) multiplets, characterized by approximately equal masses [6].

The quark model provided a coherent and satisfactory classification of mesons and baryons, but it did not yet hint at the existence of a second hadronic family. For a long time, in fact, it was believed that quarks were not real particles but merely mathematical entities. Gell-Mann himself, in his 1964 paper, wrote: "It is fun to speculate about the behavior of quarks if they were physical particles with finite mass, instead of purely mathematical entities, as they would be in the limit of infinite mass... A search for stable quarks at the most powerful particle accelerators would help confirm their non-existence as real objects."

The very idea that hadrons had fundamental constituents was definitely against the mainstream. With the increasing number of observed hadrons, the idea had developed that none of them had a special status as "elementary." Instead, it was thought that they were all com-posed of one another, according to the so-called bootstrap model, in which the properties of hadrons would be determined by self-consistency. In 1961, Geoffrey Chew, a leading advocate of this "hadron democracy" movement, argued that "The traditional association of fields with strongly interacting particles is meaningless," and that, regarding strong interactions, quantum field theory was not only sterile but, "like an old soldier, is destined not to die but simply to fade away." In this view, there was no room for quarks as elementary particles. Despite the success of the quark model, it was only in the late 1960s, thanks to deep-inelastic electron scattering experiments conducted at the Stanford Linear Accelerator Center (SLAC), that the bootstrap model was abandoned.

Furthermore, the hierarchy of quark masses was established much later. In early works, quarks were considered very heavy to explain, on kinematic grounds, why bound states were not ionized in high-energy reactions. Today, this interpretation falls apart in light of QCD, where confinement is a phenomenon that persists even in the limit of massless quarks. The first information about quark masses was obtained in the late 1960s thanks to current algebra, studying the problem of chiral symmetry breaking.

## 4  The role of weak interactions

The concept of a second family of quarks could not emerge without a connection to weak interactions. At that time, these were described through the product of two charged currents, generalizing Fermi's interaction. Using modern terminology and in the absence of "strange" hadrons, each current consisted of the sum of three terms: the first involved the electron and its antineutrino, the second the muon and its antineutrino, and the third the *up* antiquark and the *down* quark. The product of two such currents described processes such as muon decay or beta decay of nuclei. By the late 1950s, it was observed that all these processes could be described with a single coupling constant, the Fermi constant, a key property known as the universality of weak interactions: the relative weights of the three terms in the charged current were equal within the experimental errors of the time.

However, to also include weak transitions involving strange hadrons, it was necessary to add a fourth term that coupled the *up* antiquark to the *strange* quark. Experimental evidence showed that this fourth term had a weight about five times smaller than the others. In 1963, Cabibbo's theory reconciled this discrepancy with the principle of universality. Cabibbo proposed representing the weights of the two hadronic terms as the legs of a right triangle, where the hypotenuse corresponded to the weight of the leptonic term. In other words, the hadronic current could be described as a single term with the same weight as the leptonic current. In this representation, the *up* antiquark was coupled to a linear combination of the *down* quark and the *strange* quark, with coefficients given by the cosine and sine of a fundamental angle, known as the Cabibbo angle (Fig. 2). With this insight, Cabibbo not only preserved the universality of weak interactions but also made a crucial step toward the introduction of the second family of quarks [7]. The decisive breakthrough came from solving a different and seemingly unrelated problem: the probability amplitudes calculated in Fermi's theory at higher orders of perturbative expansion were found to diverge. This meant that, at energies above a certain limit, Fermi's theory had to be replaced by a new one. For a long time, it was believed that this limit was on the order of a few hundred GeV, a value effectively unreachable with the technology of the time. However, in 1967, an important study by Ioffe and Shabalin demonstrated that this limit was actually much lower, on the order of a few GeV, undermining the reliability of the theory at the very energies typical of the processes being studied. Moreover, due to these divergences, the theory predicted probabilities for certain processes that were in clear conflict with experimental data.

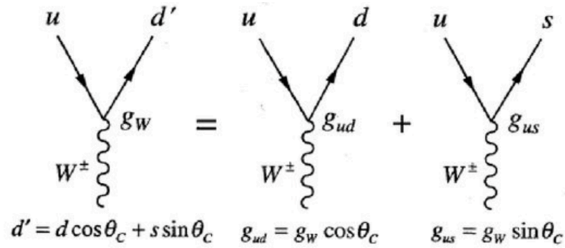

**Figure 2:** Coupling of the antiquark *up* to quark *down* and the quark *strange* according to the Cabibbo theory.

The most promising solution to eliminate these divergences involved introducing a charged vector boson as the mediator of interactions. However, even this hypothesis was not sufficient to resolve all divergences. For instance, in Cabibbo's theory, the amplitude responsible for the mass difference between the $K_L$ and $K_S$ mesons, or the one describing the decay $K_L \to \mu^+ \mu^-$, exhibited divergent contributions due to the exchange of two virtual charged bosons. In 1970, Glashow, Iliopoulos, and Maiani (GIM) identified a mechanism to eliminate this class of divergences by modifying Cabibbo's theory with the introduction of a fourth quark, called the *charm* quark, with an electric charge of +2/3. The charged current thus acquired a new term, in which the *charm* antiquark coupled to a combination of the *down* and *strange* quarks orthogonal to the one introduced by Cabibbo. This generated new Feynman diagrams, involving the exchange of the *charm* quark, which were added to the existing diagrams with the exchange of the *up* quark. Thanks to the orthogonality of the couplings, if the *up* and *charm* quarks had the same mass, the diagrams would have canceled each other out exactly (Fig. 3). In reality, a mass difference of just a few GeV between the two quarks was sufficient to eliminate the divergent quantities and reconcile theoretical predictions with experimental data. Furthermore, the weak interactions involving the new quark preserved universality through the same angle introduced by Cabibbo. The GIM mechanism, named after its authors, represents one of the cornerstones of electroweak theory and is fully integrated into the Standard Model of fundamental interactions. This mechanism is essential for understanding the suppression of flavor-changing neutral currents.

# 5 The November Revolution

Four years after the hypothesis of the *charm* quark, on November 11, 1974, the discovery of a particle with a mass of approximately 3 GeV was announced by two independent experiments. At the Brookhaven National Laboratory, Samuel Ting's group observed it in proton-fixed-target collisions, while at SLAC, Burton Richter's group identified it in electron-positron collisions. The particle, called J/$\psi$, is a bound state formed by a *charm* quark and its antiquark. The decay of the J/$\psi$ into pairs of mesons containing *charm* quarks is prohibited by kinematics, while its decay into other mesons requires the contribution of strong interactions in a weak coupling regime. Despite having a mass typical of a hadron, the particle is distinguished by an unusually long lifetime. This discovery confirmed the predictions of the GIM mechanism and definitively identified the family structure of the Standard Model.

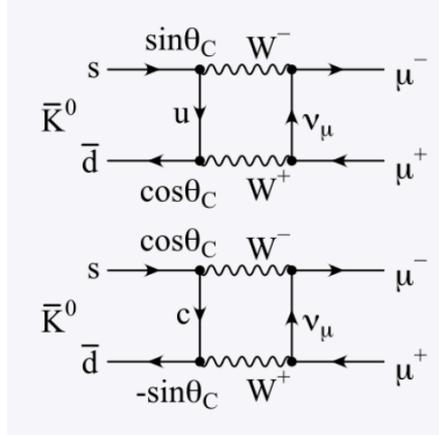

**Figure 3:** Amplitude cancellation for the $\bar{K}^0 \to \mu^+\mu^-$ transition by the GIM mechanism.

Italian physics played a leading role in the so-called "November Revolution." The discovery of the J/$\psi$ could have been entirely Italian. The first electron-positron collider, AdA (anello di accumulazione), was conceived by Bruno Touschek and built in Frascati in 1960 [8]. Its successor, ADONE, completed in 1967 and operational from 1969, with important contributions from Raul Gatto and a very young Nicola Cabibbo, was capable of reaching a center-of-mass energy of 3 GeV. It could have anticipated the American groups in the discovery of the J/$\psi$, but had to settle for confirming it on November 13, 1974.

With this discovery, the quark sector aligned perfectly with the lepton sector. There are two quarks with a charge of -1/3 (*down* and *strange*) and two with a charge of +2/3 (*up* and *charm*). The *up* and *down* quarks describe ordinary hadrons, while *charm* and *strange* form the second family, associated with new hadronic states. Completing the theoretical framework, an important work by Bouchiat, Iliopoulos, and Meyer in 1972 demonstrated that electroweak interactions are consistent only in the presence of entire families that replicate the quantum numbers of *up*, *down*, electron, and neutrino. There are thus no theoretical restrictions on the existence of additional quarks and leptons, provided they are organized into complete families.

# 6  The third family arrives

The inclusion of a new family of fermions was motivated by the CP violation in the decays of K mesons. In 1973, Kobayashi and Maskawa proposed a mechanism to explain CP symmetry violation, hypothesizing the need for at least three generations of quarks. Four years later, in 1977, studying proton-antiproton collisions at Fermilab, a group led by Leon Lederman discovered the ϒ particle, a meson composed of a quark with a charge of -1/3 and its antiquark. This quark, called *bottom*, was found to have a much larger mass than the quarks with the same charge from the first two generations. The discovery of its +2/3 charged partner, the *top* quark, required many more years. It was only in 1995, thanks to the CDF and D0 experiments at Fermilab, that the *top* quark was finally identified. Its long-awaited revelation was made possible through the study of proton-antiproton collisions at extremely high energies, necessary to produce such a massive particle. The third family of quarks was later joined by the lepton family, composed of the tau lepton, discovered at SLAC in 1975, and the corresponding tau neutrino, observed for the first time only in 2000 in the DONUT experiment at Fermilab. Finally, with the discovery of neutrino oscillations and their detailed study, the new millennium has clarified the structure of mixing in the lepton sector, revealing a completely different picture compared to the hadronic sector.

How many families of particles exist in nature? And could there be room for families that have not yet been observed? By studying the properties of the Z vector boson and its invisible decays, we know that there are no more than three light neutrinos. The known neutrinos have extremely small masses, more than a million times smaller than that of the electron, making the existence of a fourth neutrino highly unlikely. To evade the limit set by Z decays, such a neutrino would have to be extremely massive. Furthermore, a number of light neutrinos different from three would be incompatible with the evolution of the universe. It would alter both the primordial abundance of light elements and the properties of the cosmic microwave background radiation. Similarly, the existence of a fourth generation of fermions would significantly change the production and decay probabilities of the Higgs boson, which are now measured with enough precision to exclude the existence of new quarks and leptons with the same characteristics as the first three generations.

# 7 In search of an organizing principle

With upcoming experiments, this overall picture could be enriched by unexpected new results, potentially emerging from other areas of research. The problem of families and fermionic masses is closely tied to the origin of the electroweak scale and its stability relative to the Planck scale. Valuable insights into this issue could come from both existing accelerators and those currently being designed. As we will see, many approaches to explaining the origin of families predict, to some extent, an increase in neutral current processes with flavor-changing interactions compared to the predictions of the GIM mechanism. Numerous experiments aimed at measuring extremely rare processes in the Standard Model are currently underway or in the planning stages. Moreover, most of the matter that constitutes the universe exists in the form of dark matter, whose nature remains unknown. If dark matter is linked to a yet unidentified particle, there could be a direct connection to the origin of families and their interactions. The search for dark matter is one of the leading developments in particle physics and astrophysics today.

Although significant future developments cannot be ruled out, detailed measurements of fermionic masses and mixing angles have so far provided a coherent general picture, sufficient to raise many fundamental questions. Why are there exactly three families? What determines the hierarchy of charged fermion masses, with the third generation being much more massive than the second, and the second much more massive than the first (Fig. 4)? Why are the mixing angles between quarks so small? And why do the leptonic mixing angles not follow the same pattern observed in the quark sector? In the Standard Model, every mass, mixing angle, and observable phase corresponds to a free parameter, which by definition cannot be calculated within the framework of the model itself. The theory allows, in principle, any value for these parameters. It is important to keep in mind that reducing these parameters to a more economical set of variables may not be possible and could even represent a false problem. The history of physics provides instructive examples in this regard. A famous case is the proportions of planetary orbits in the solar system, which were long considered a fundamental enigma. Kepler himself proposed a solution to this problem in his work Harmonices Mundi. Today, however, we believe that such proportions do not reflect a fundamental property of nature, given the non-privileged role of our planetary system in the universe. Similarly, a comparable situation could apply to the parameters that describe fermion masses in the Standard Model. If, however, one accepts the idea that such parameters could be calculable, it becomes necessary to go beyond the Standard Model [9, 10]. Below, we will review attempts to calculate the masses of quarks and leptons in terms of a smaller number of independent parameters.

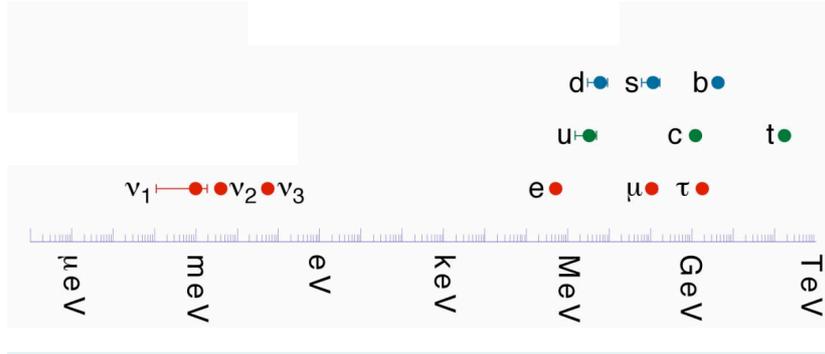

**Figure 4:** Fermion masses, logarithmic scale. A GeV is approximately the mass of the proton.

# 8 Some hints

A first attempt to reduce the number of parameters draws inspiration from the origin of electroweak and strong interactions. These interactions could be the low-energy imprint of a unified theory with a symmetry that fully manifests only at very high energies, far beyond those accessible today. It is hard not to be struck by the fact that the quantum numbers of quarks and leptons, which are inexplicable within the context of the Standard Model, find a natural explanation in the simplest multiplets of grand unification theories. Although there is currently no experimental confirmation of this proposal, it is plausible that it could play a crucial role in understanding fundamental interactions. In this context, the distinction between quarks and leptons loses significance, and one would expect relationships between the parameters involving them. Minimal models, for example, predict that the masses of charged leptons are approximately equal to those of quarks with a charge of -1/3. However, to make this relationship compatible with experimental data, corrections and additional parameters must be introduced, at the expense of calculability understood as a reduction in independent parameters. Nonetheless, approximate relationships of this type could provide valuable insights into the structure of fermion masses.

Since the third-generation fermions have significantly larger masses than those of the first two generations, it is plausible that the smaller masses are generated through radiative corrections originating from the larger ones - an idea favored by Steven Weinberg, who devoted one of his final works to it [11]. For example, one could imagine a scenario in which, at the first order of perturbative expansion, only the third-generation fermions have non-zero masses. However, within the Standard Model, it is not possible to generate the masses of the remaining fermions through radiative effects. In fact, the starting point enjoys a chiral symmetry that makes it stable with respect to any perturbative correction. We must, therefore, move beyond the Standard Model and start from a theory where the fermions of the first two generations are massless due to accidental reasons, not attributable to symmetry. This theory would contain new fields and interactions that, by violating chiral symmetry, could generate the masses of the first two generations of fermions through quantum corrections. However, this scenario comes at a significant cost. In addition to the great variety of possible microscopic realizations, the main problem is the number of independent parameters required to describe the masses and interactions of the new states. These parameters must be finely tuned to reproduce the fermion masses and mixing angles while respecting the constraints imposed by experiments on flavor-changing neutral current processes. Thus, the original problem is not only unresolved but is further amplified. Moreover, this approach does not provide an explanation for why there are only three generations.

# 9   Quarks and leptons as composite states

A possible explanation for the replication of generations is that the Standard Model fermions are, in reality, composed of more fundamental particles. Similar to how hydrogen, deuterium, and tritium consist of a proton and, respectively, zero, one, and two neutrons (Fig. 5), quarks and leptons might also have a substructure based on more elementary components. While fascinating, this idea faces significant challenges. Experiments probing the internal structure of quarks and leptons indicate that the spatial size $r$ of a hypothetical composite state would have to be extremely small, many orders of magnitude smaller than the Compton wavelength of the most massive fermion. This result can be rephrased in terms of mass scales: using the compositeness scale $\Lambda = 1/r$ instead of the spatial size $r$, it follows that the mass $m$ of quarks and leptons would be much smaller than $\Lambda$. Such a situation is unprecedented because, in known composite states, the typical mass is equal to or greater than the compositeness scale. For example, in the hydrogen atom, the reduced mass of the system is about two orders of magnitude greater than the inverse of the Bohr radius. For quarks and leptons, the opposite would occur, a feature that challenges our current understanding. This hypothesis seems justified only within a theory where, to a first approximation, quarks and leptons have zero mass, with small corrections relative to the scale $\Lambda$. In general, these corrections depend on both the electroweak scale and the scale $\Lambda$, and their inclusion entails several difficulties. The result must ensure the suppression of flavor-changing neutral currents, protected by the GIM mechanism in the Standard Model, and the suppression of processes involving baryon number violation. The amplitudes of such processes are typically suppressed by inverse powers of the scale Λ, which would need to be pushed many orders of magnitude above the energies currently explored. Confining the known fermions within such small spatial dimensions also requires a regime of strong interaction, making quantitative calculations in these theories often prohibitive. So far, attempts in this direction have not advanced the calculability of physical parameters [12].

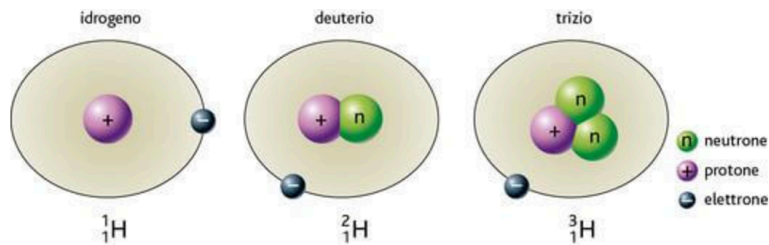

**Figure 5:** Isotopes of the hydrogen atom.

However, the idea of compositeness has gained renewed interest in a different context, envisioning that Standard Model fermions acquire mass through mixing with composite states. The hypothesis is that these are produced by a strongly interacting sector at the TeV scale, motivated by efforts to explain the disparity between the electroweak scale and the Planck scale. The stronger the mixing between a fermion and the composite states, the greater its degree of compositeness and the higher its mass. Strongly composite fermions are heavier, while weakly composite ones are lighter. This dynamic allows for a "dual" description, assuming that the fermions are represented by a wavefunction localized along an interval of a new spatial dimension, distinct from the three known ones and currently inaccessible. In this description, the Higgs field is localized at one end of the interval and provides a common mass scale, the electroweak scale, to all particles. Fermions with wavefunctions close to (or far from) the end where the Higgs boson is located turn out to be heavy (or light). The degree of compositeness of a fermion is thus proportional to the overlap of its wavefunction with the Higgs boson's position (Fig. 6) [13]. Unfortunately, in its simplest form, this idea does not guarantee sufficient suppression of flavor-changing neutral currents in the presence of new physics at the TeV scale. Although the realistic version of this approach involves a high number of independent parameters and does not provide advantages in terms of calculability, it offers a geometric reinterpretation of the family problem. Perhaps the masses and mixing angles of fermions are linked to the localization of particles in a new spatial dimension and the geometry of that dimension. This is precisely what is pre-dicted by string theory, the leading candidate for unifying all known interactions, including gravity. A coherent formulation of this theory requires a spacetime with ten dimensions. To reconcile this feature with observations, six spatial dimensions must be rendered inaccessible by compacting them into extremely small extensions, requiring extraordinarily high energies to explore. In this context, quarks and leptons are variously localized within the six compact dimensions, providing the elements necessary to interpret masses and mixing angles from a geometric perspective. Furthermore, the problem of the replication of generations finds an explanation in the very geometry of the compact space, which might only allow for the three observed families.

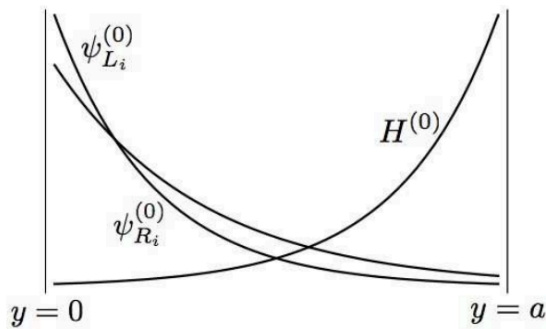

**Figure 6:** Wave functions for the left (L) and right (R) components of a fermion and their superposition with the Higgs wave function (H) in an extra spatial dimension.

This is a fascinating hypothesis, but it faces formidable challenges. There are countless ways to realize the compact space, with no clear preference for one over the others. Could our universe correspond to one of these solutions? If so, which one? Even with partial information, the masses and mixing angles of fermions depend on the values taken by a set of initially undetermined variables, the so-called moduli of the theory, which describe the shape and size of the compact space. Although string theory contains the ingredients to calculate the parameters of the Standard Model, this calculation remains beyond our reach for now. Additionally, if all possible compactifications were dynamically equivalent, we would face a situation similar to that of ancient astronomers, who could observe a single solar system without knowing it was just one among many in the universe.

## 10 Symmetry to the rescue?

A common feature of nearly all these scenarios is the presence of a symmetry acting in the space of generations, emerging when small effects are neglected. Regardless of the specific microscopic realization, this "flavor" symmetry could guide us toward a more fundamental framework, much like the Eightfold Way led to the quark model of hadrons. However, compared to hadrons, identifying the correct symmetry and the multiplets to which quarks and leptons are assigned is far more complex, as there are many plausible choices. For instance, one might start from the observation that the first two generations of charged fermions are much lighter than the third. Neglecting the masses of the first two generations, the theory acquires a chiral symmetry, which is then broken by small parameters to achieve realistic results. An equally reasonable observation is that the observed hierarchies in mass ratios and mixing angles can be approximated by powers of a small parameter, on the order of the Cabibbo angle. If this quantity is interpreted as the breaking parameter of a U(1) symmetry, the required powers can be reproduced by assigning appropriate charges to fermions of different generations. There are countless examples of this kind, each with intriguing aspects, but none emerge as more convincing than the others. Most of them use a large number of independent parameters and require a dedicated symmetry-breaking sector, often characterized by significant complexity. Moreover, they fail to explain why there are exactly three families. The primary utility of this approach perhaps lies in models with a scale of new physics close to energies accessible today. These models generally predict rare transitions that, in the absence of approximate flavor symmetries, would conflict with experimental data.

The absence of clear indications in experimental data regarding the type of symmetry and the associated breaking sector has led researchers to seek answers within the framework of string theory. This theory predicts both diffeomorphism invariance, which underpins general relativity, and the local symmetries on which the Standard Model is based. But are there indications of flavor symmetries? In some concrete cases analyzed so far, such symmetries seem to emerge from the properties of the compact space in which the six additional dimensions are confined. Once the compact space is fixed, quarks and leptons are uniquely organized into multiplets of a specific flavor symmetry [14]. A distinctive feature of this framework is the interplay between transformations acting on quarks and leptons and those involving the moduli - the variables that describe the size and shape of the compact space [15]. Not all symmetry groups permit this realization, and the corresponding constructions, guided by the selective rules imposed by string theory, appear less arbitrary. However, it is still too early to determine whether this direction can provide a satisfactory answer to our questions.

Nearly ninety years after the discovery of the muon by Neddermeyer and Anderson, the mystery of the families remains unsolved. Every attempt to identify a principle capable of explaining the observations seems destined to fail. It is as if we have pieces of a gigantic puzzle. We have assembled some initial parts, but we cannot determine whether the remaining pieces can complete the picture. Perhaps among the available pieces, there are extraneous ones that should be discarded. Or, more simply, we are still missing the overall vision, the complete picture that could provide the logic necessary to finish the work.

# Acknowledgments

I am deeply grateful to Carlo Broggini, Luciano Maiani, and Fabio Zwirner for their valuable comments and suggestions during the preparation of this text.